\documentclass[12pt,a4,portrait]{article}
\usepackage{latexsym}
\usepackage{hyperref}
\usepackage{graphicx}

\newcommand{\be}{\begin{equation}}
\newcommand{\ee}{\end{equation}}
\newcommand{\beq}{\begin{eqnarray}}
\newcommand{\eeq}{\end{eqnarray}}
\newcommand{\bed}{\begin{displaymath}}
\newcommand{\eed}{\end{displaymath}}
\newcommand{\bc}{\begin{center}}
\newcommand{\ec}{\end{center}}
\newcommand{\bi}{\begin{itemize}}
\newcommand{\ei}{\end{itemize}}
\newcommand{\bn}{\begin{enumerate}}
\newcommand{\en}{\end{enumerate}}

\newcommand{\rw}{ {\rm w} }
\newcommand{\dd}{ {\rm d} }
\newcommand{\pmp}{ \matrix{ + \vspace{-2mm}\cr \mbox{\scriptsize(} \! - \! \mbox{\scriptsize)}} }
\newcommand{\mpp}{ \matrix{ - \vspace{-2mm}\cr \mbox{\scriptsize(} \! + \! \mbox{\scriptsize)}} }
\newcommand{\mxp}{ \matrix{ - \vspace{-2mm}\cr \mbox{\scriptsize\phantom(} \! \phantom+ \! \mbox{\scriptsize\phantom)} } }
\newcommand{\xpp}{ \matrix{ \phantom- \vspace{-2mm}\cr \mbox{\scriptsize(} \! + \! \mbox{\scriptsize)}} }

\begin{document}
\baselineskip 1.2\baselineskip
\thispagestyle{empty}

{\phantom .}

\begin{center}

%\vspace{3.0cm}

{\Large\bf{Scalar-tensor cosmologies: general relativity \linebreak as a fixed %\linebreak
point of the Jordan frame scalar field}}

\vspace{2.0 cm}

Piret Kuusk$^{1 a}$, Laur J\"arv $^{2 a}$ and Margus Saal$^{3  b}$

\vspace{0.3cm}
{\it {\phantom x}$^{a}$ Institute of Physics, University of Tartu,
Riia 142, Tartu 51014, Estonia \linebreak
{\phantom x}$^{b}$ Tartu Observatory, T\~oravere 61602, Estonia}

\end{center}

\vspace{1,5cm}

\begin{abstract}
We study the evolution of homogeneous and isotropic, flat cosmological
models within the general
scalar-tensor theory of gravity with arbitrary coupling function and
potential and scrutinize its limit to general relativity. 
Using the methods
of dynamical systems for the decoupled equation of the Jordan frame scalar
field we find the fixed
points of flows in two cases: potential domination and matter domination.
We present the conditions on the mathematical form of the coupling function
and potential which
determine the nature of the fixed points (attractor or other).
There are two types of fixed points, both are characterized by cosmological
evolution mimicking
general relativity, but only one of the types is compatible with the Solar
System PPN constraints.
\end{abstract}

\vspace{1cm}
{\bf PACS}: 98.80.Jk, 04.50.+h, 04.40.Nr

\vspace{5 cm}
$^1$ Electronic address: piret@fi.tartu.ee

$^2$ Electronic address: laur@fi.tartu.ee

$^3$ Electronic address: margus@fi.tartu.ee

\newpage

\section{Introduction}

Scalar-tensor theories (STT) of gravitation \cite{will, fm,faraoni}
have emerged in different contexts of theoretical physics, e.g.
in Kaluza-Klein type unifications, supergravity, and
low energy approximations of string theories.
In cosmology STT has been invoked to model the accelerated expansion of
inflation 
and dark energy. 
However, observations in the Solar System tend to indicate that in the
intermediate-range distances the present Universe around us is
successfully described by
the Einstein tensorial gravity alone
\cite{tests1, tests2, tests3}.
This means that only such models of scalar-tensor gravity are viable which
in their late time cosmological evolution
imply local observational consequences very close to those of Einstein's
general relativity (GR) \cite{dn}.

The methods of dynamical systems provide natural
tools to analyze the problem.
In this paper, summarizing our recent work \cite{meie4}, 
we take the Jordan frame and consider general
scalar-tensor theories which
contain two functional degrees of freedom, the coupling function
$\omega (\Psi)$ and the scalar potential $V(\Psi)$.
We perform the dynamical systems
analysis for the flat Friedmann-Lema\^{\i}tre-Robertson-Walker (FLRW)
backgrounds with ideal barotropic fluid matter.
Our strategy is to find the fixed points for
the scalar field dynamics and compare these
with the conditions of the limit of general relativity in the Solar System,
as established by the parameterized post-Newtonian (PPN) formalism.
Therefore, if the functional forms of $\omega (\Psi)$ and $V(\Psi)$
are specified from some considerations (e.g. the compactification manifold),
our results allow to determine the fixed points along with their type and
thus
immediately decide whether general relativity is an attractor,
i.e. whether the model at hand is viable or not.

%%%%%%%%%%%%%%%%%%%%%%%%%%%%%%%%%%%%%%%%%%%%%%%%%%%%%%%%%%%%
%%%%%%%%%%%%%%%%%%%%%%%%%%%%%%%%%%%%%%%%%%%%%%%%%%%%%%%%%%%%%%%

\section{Scalar-tensor cosmology as a dynamical system and the limit of general relativity}

We consider  a general scalar-tensor theory
in the Jordan frame given by the action functional
    \beq \label{jf4da}
S  = \frac{1}{2 \kappa^2} \int d^4 x \sqrt{-g}
                    \left[ \Psi R(g) - \frac{\omega (\Psi ) }{\Psi}
                \nabla^{\rho}\Psi \nabla_{\rho}\Psi
                  - 2 \kappa^2 V(\Psi)  \right]
                   + S_{m}(g_{\mu\nu}, \chi_m) \,.
\eeq
Here $\omega(\Psi)$ is a coupling function and $V(\Psi)$ is a
scalar potential, $\nabla_{\mu}$
denotes the covariant derivative with respect to the metric
$g_{\mu\nu}$ and
$S_{m}$ is the matter part of the action
as all other fields are included in $\chi_m$.
In order to keep the effective Newtonian gravitational constant 
positive \cite{nordtvedt70} we assume that $0 < \Psi < \infty$.

The field equations for the flat ($k=0$)
FLRW line element and perfect barotropic fluid
matter, $p=\rw \rho$, 
read
\beq
\label{00}
H^2 &=&
- H \frac{\dot \Psi}{\Psi}
+ \frac{1}{6} \frac{\dot \Psi^2}{\Psi^2} \ \omega(\Psi)
+ \frac{\kappa^2}{3} \frac{ \rho}{\Psi}
+ \frac{\kappa^2}{3} \frac{V(\Psi)}{\Psi} \,,
\\ \nonumber \\
\label{mn}
2 \dot{ H} + 3 H^2 &=&
- 2 H \frac{\dot{\Psi}}{\Psi}
- \frac{1}{2} \frac{\dot{\Psi}^2}{\Psi^2} \ \omega(\Psi)
- \frac{\ddot{\Psi}}{\Psi}
- \frac{\kappa^2}{\Psi} \rw \rho
+ \frac{\kappa^2}{\Psi} \ V(\Psi) \,,
\\ \nonumber \\
\label{deq}
\ddot \Psi &= &
- 3H \dot \Psi
- \frac{1}{2\omega(\Psi) + 3} \ \frac{d \omega(\Psi)}{d \Psi} \  \dot
{\Psi}^2
+ \frac{\kappa^2}{2 \omega(\Psi) + 3} \ (1-3 \rw) \ \rho  \nonumber \\
& &\qquad + \frac{2 \kappa^2}{2 \omega(\Psi) + 3} \ \left[ 2V(\Psi) - \Psi \
\frac{d V(\Psi)}{d \Psi}\right] \, ,
\\ \nonumber \\
  \label{matter_conservation}
\dot{\rho} &=& - 3 H \ (\rw+1) \ \rho  \, .
\eeq
Here $H \equiv \dot{a} / a$
and we assume $\rho \geq 0$.
Eqs. (\ref{00})--(\ref{matter_conservation})
are too cumbersome to be solved analytically,
but useful information about the general characteristics of solutions
can be obtained by rewriting (\ref{00})--(\ref{matter_conservation})
in the form of
a dynamical system and finding
the fixed points which describe the asymptotic behaviour of solutions.

The phase space of the system
is spanned by four variables $\{ \Psi, \dot{\Psi}, H, \rho \}$.
Defining $\Psi \equiv x, \dot{\Psi} \equiv y$ the dynamical system
corresponding to
equations
(\ref{00})-(\ref{matter_conservation}) can be written
as follows:
\beq
\label{general_dynsys_x}
\dot{x} &=& y \,, \\
\label{general_dynsys_y}
\dot{y} &=& -\frac{1}{2\omega(x)+3} \! \left[
\frac{d\omega(x)}{dx} y^2 \! -\kappa^2 \! \! \left( 1-3\rw \right)\rho
+2 \kappa^2 \left( \frac{dV(x)}{dx}x-2V(x) \right)
\right] \! - 3Hy , \\
\label{general_dynsys_H}
\dot{H} &=& \frac{1}{2x(2\omega(x)+3)} \left[
\frac{d\omega(x)}{dx} y^2 -\kappa^2 \left( 1-3\rw \right)\rho
+2 \kappa^2 \left( \frac{dV(x)}{dx}x -2V(x) \right)
\right] \nonumber \\
&& \quad -\frac{1}{2x} \left[ 6H^2x + 2Hy - \kappa^2 (1-\rw)\rho -2\kappa^2
V(x)
\right] \,, \\
\dot{\rho} &=& -3 H (1+\rw) \rho \,.
\label{general_dynsys_rho}
\eeq

Based on these equations we may
make a couple of quick qualitative observations about
some general features of the solutions.
For example, the limit $\Psi \rightarrow 0$ in general implies
$|\dot{H}| \rightarrow \infty$,
hence the solutions can not safely pass from positive to negative values of
$\Psi$
(from ``attractive'' to ``repulsive'' gravity),
but hit a space-time singularity as the curvature invariants diverge.
Similarly, the limit $2\omega + 3 \rightarrow 0$ implies
$|\dot{H}| \rightarrow \infty$ with the same conclusion that
passing through $\omega(\Psi)=-\frac{3}{2}$
would entail a space-time singularity and is impossible.
(Let us remark here, that these observations are quite general and do not
preclude specially fine-tuned solutions in some fine-tuned models which may
remain regular while crossing these points \cite{gunzing_2001}.)

The limit $\frac{1}{2\omega + 3} \rightarrow 0$ deserves a more
closer examination. Let us define $x_{\star}$ by
$(2\omega(x_{\star}) + 3)^{-1} =0.$
Expressing $H$ from the Friedmann constraint (\ref{00}),
\be
\label{H_constraint}
H = -\frac{y}{2 x}\pmp 
%\begin{matrix}
%+ \vspace{-2mm} \cr
%  \mbox{\scriptsize(} \! - \! \mbox{\scriptsize)}
%\end{matrix}
\sqrt{\left( 2\omega(x)+3 \right) \frac{y^2}{12x^2}
+ \frac{\kappa^2 (\rho + V(x))}{3x} } \, ,
\ee
makes clear that $|H|$ diverges as $x \to x_{\star}$,
unless also $y \to 0$ at the same time.
What happens in the latter case is
determined by the first term under the square root above. We can
compute its limit by Taylor expanding
\beq
&& \!\!\!\!\!\!\! %\qquad  \qquad \qquad \qquad \qquad 
\lim_{y \to 0 \atop x \to x_{\star}} (2\omega(x)+3)y^2  =
%\lim_{\Delta y \to 0 \atop \Delta x \to 0}
%\frac{y^2|_{y=0} + 2 y|_{y=0} \Delta y + (\Delta y)^2}
%{\frac{1}{2\omega(x)+3}\Big|_{x=x_{\star}}
%+ \frac{ \dd}{\dd
%x}\left(\frac{1}{2\omega(x)+3}\right)\Big|_{x=x_{\star}}\Delta x
%+ \frac{1}{2} \frac{\dd^2}{\dd
%x^2}\left(\frac{1}{2\omega(x)+3}\right)\Big|_{x=x_{\star}} (\Delta x)^2 +
%\dots  }
\nonumber \\
%\hspace*{-5mm}
&&\quad 
= \lim_{r \to 0} \frac{r^2 \sin^2\theta}
{\frac{\dd}{\dd x}\left(\frac{1}{2\omega(x)+3}\right)\Big|_{x=x_{\star}} r
\cos\theta
+ \frac{1}{2}
\frac{d^2}{dx^2}\left(\frac{1}{2\omega(x)+3}\right)\Big|_{x=x_{\star}} r^2
\cos^2\theta + \dots} \nonumber \\
\vspace{15mm}
&&
\left \{ \begin{array}{ll}
= 0 \, , & $if$ \quad \frac{\dd}{\dd
x}\left(\frac{1}{2\omega(x)+3}\right)\Big|_{x=x_{\star}} \neq 0 \,,
\vspace{2mm}\\
\sim  \tan^2 \theta \, , &  $if$ \quad  \frac{\dd}{\dd
x}\left(\frac{1}{2\omega(x)+3}\right)\Big|_{x=x_{\star}}=0 \; , \quad
\frac{\dd^2}{\dd x^2}\left(\frac{1}{2\omega(x)+3}\right)\Big|_{x=x_{\star}}
\neq 0 \,, \vspace{2mm}\\
= \infty \, , & $if$ \quad  \frac{\dd}{\dd
x}\left(\frac{1}{2\omega(x)+3}\right)\Big|_{x=x_{\star}}= 0 \;, \quad
\frac{\dd^2}{\dd x^2}\left(\frac{1}{2\omega(x)+3}\right)\Big|_{x=x_{\star}}
= 0 \,,
\end{array}
\right.
\label{taylor_expansion}
\eeq
where $\Delta x = r \cos\theta$, $\Delta y = r \sin\theta$ was taken.
(We have neglected the unphysical direction $|\theta| = \frac{\pi}{2}$
that corresponds
to approaching the point $(x=x_{\star},y=0)$
along the line $x=x_{\star}$ where
$|H|$ is divergent.)
So, in the limit $x \to x_{\star}, y \to 0$
the value of $H$ is determined by the lowest non-zero derivative of
$\frac{1}{2\omega(x) + 3}$.
If the both the first and second derivative vanish,
then $|H|$ diverges implying
a spacetime singularity.
If the first derivative vanishes but the second derivative is not zero,
\be
\frac{\dd}{\dd x}\left(\frac{1}{2\omega(x)+3}\right)\Big|_{x=x_{\star}}=0 \;
, \qquad  \frac{\dd^2}{\dd
x^2}\left(\frac{1}{2\omega(x)+3}\right)\Big|_{x=x_{\star}} \neq 0 \,,
\ee
then $H$ is finite but (possibly) different for each solution as it
depends on the angle of approach $\theta$, while
the Friedmann equation in this case acquires an extra term
when compared to general relativity.
If the first derivative is not zero,
\be \label{first_derivative_nonzero}
\frac{\dd}{\dd x}\left(\frac{1}{2\omega(x)+3}\right)\Bigg|_{x=x_{\star}}=
\frac{1}{(2\omega(x_{\star})+3)^2}\frac{\dd \omega}{\dd
x}\Bigg|_{x=x_{\star}} \neq 0 \,,
\ee
then $H$ approaches the value
$H_{\star}^2 = \frac{\kappa^2}{3 x_{\star}}(\rho+V(x_{\star}))$,
mimicking the Friedmann equation of general relativity
with $8 \pi G = \frac{\kappa^2}{x_{\star}}$ and
$\Lambda = \frac{\kappa^2}{x_{\star}}V(x_{\star})$.

To summarize, we have just observed that in the limit
(a) $\frac{1}{2\omega(x) + 3} \rightarrow 0$,
(b) $y \rightarrow 0$
the Friedmann constraint (\ref{H_constraint}) tends to the form
of general relativity
if
(c) $\frac{1}{(2\omega(x_{\star})+3)^2}\frac{\dd \omega}{\dd
x}\Big|_{x=x_{\star}} \neq 0$.
It must be also emphasized here
that the process of taking the Taylor expansion (\ref{taylor_expansion})
hinges on the assumption that (d)
$\frac{1}{2\omega(x)+3}$ is differentiable (derivatives do not diverge)
at $x_{\star}$.
In this context one may also ask
when the full set of equations
(\ref{general_dynsys_x})-(\ref{general_dynsys_rho})
attains the form
of general relativity. It is easy to see that besides
(a)-(d) one must also impose
\be
\frac{1}{2\omega(x)+3} \frac{\dd \omega}{\dd x} y^2 =
\frac{1}{(2\omega(x_{\star})+3)^2}\frac{\dd \omega}{\dd x}
\left( (2\omega(x)+3) y^2 \right) \to 0 \,,
\ee
but the latter is automatically satisfied if (c) holds, due to
(\ref{taylor_expansion}), (\ref{first_derivative_nonzero}).
Therefore we may tentatively call the conditions (a)-(d)
`the general relativity limit of scalar-tensor flat FLRW
cosmology'.

It is interesting to compare the cosmological GR limit to the GR limit
obtained from PPN, which
characterizes the slow motion approximation in a centrally symmetric
gravitational field. Although the mathematical assumptions underlying the
PPN formalism are clearly different from our
cosmological reasoning above, we may still ask whether the results of both
schemes agree with each other.
In the context of PPN it is well established that
the solutions of scalar-tensor theory approach those of general relativity
when \cite{nordtvedt70}
\be \label{PPN_condition}
\frac{1}{2\omega(x)+3} \to 0 \,, \qquad
\frac{1}{(2\omega(x)+3)^3} \frac{\dd \omega}{\dd x} \to 0 \,.
\ee
Comparison shows that
the cosmological conditions (a)-(d) are marginally stricter than the PPN
condition
(\ref{PPN_condition}), since (a), (c), (d) imply that (\ref{PPN_condition})
is satisfied, but (\ref{PPN_condition}) does not necessarily guarantee that
(c) or (d) holds.

Let us also note that there is also another special case
$x_{\bullet}$, realized at
\be \label{second_GR_limit}
\rho = 0, \qquad
y = 0, \qquad
2V(x_{\bullet}) - x_{\bullet} \ \frac{\dd V(x)}{\dd x}\Big|_{x_{\bullet}} =0
\,,
\ee
when the cosmological equations
(\ref{general_dynsys_x})-(\ref{general_dynsys_rho})
relax to those of general relativity featuring de Sitter evolution.
However, as the value of
$\omega(x_{\bullet})$ is not fixed by the condition (\ref{second_GR_limit}),
this case does not conform with the GR limit of PPN.
Therefore, even when the limits (a)-(d) and
(\ref{second_GR_limit}) can be cosmologically indistinguishable,  Solar
System observations in the PPN framework can in principle reveal
which of the two is actually realized.
(In this paper when using the phrase `the GR limit of STT' we mean the
conditions (a)-(d),
as these take the STT cosmological equations to those of  general relativity
and also guarantee that the PPN condition is satisfied.
But note that some authors \cite{mn, carloni_2008}
have not necessarily used the same definition.)

The general relativity limit of STT is
purely given in terms of $x$ and $y$. In the following we extract from the
full
dynamical system (\ref{general_dynsys_x})-(\ref{general_dynsys_rho})
an independent subsystem for $\{ x,y \}$, find its fixed points and
check whether the limit of general relativity matches to an attractive fixed
point.

%%%%%%%%%%%%%%%%%%%%%%%%%%%%%%%%%%%%%%%%%%%%%%%%%%%%%%%%%%%%%%%%%%%%%%
%%%%%%%%%%%%%%%%%%%%%%%%%%%%%%%%%%%%%%%%%%%%%%%%%%%%%%%%%%%%%%%%%%%%%

\section{Fixed points for potential domination 
($V \not\equiv 0$, $\rho \equiv 0$)}

In the four phase space dimensions  $\{ \Psi \equiv x, \dot{\Psi} \equiv y,
H, \rho \}$
the physical trajectories
(orbits of solutions) are those
which satisfy the Friedmann constraint (\ref{00}). 
In the limit of vanishing matter density
the phase space shrinks to three dimensions
$\{x \equiv \Psi, y \equiv \dot\Psi, H \}$, where the
Friedmann constraint restrains the physical trajectories to span
two dimensions.
We may solve the Friedmann constraint for $H$, as (\ref{H_constraint}),
substitute it into Eq. (\ref{general_dynsys_y}), and thus in effect reduce
the system 2-dimensional:
\beq \label{V_dynsys}
\left \{ \begin{array}{rcl}
\dot{x} & = & y  \\
\dot{y} & = &
\left( \frac{3}{2x} - \frac{1}{2 \omega(x) + 3}\, \frac{d \omega}{dx}
\right) \, y^2
\mpp 
\frac{1}{2x} \sqrt{3 (2 \omega(x) + 3) y^2 + 12 \kappa^2 x V(x)}\ y
\\
%&& \qquad  \qquad \qquad 
& &+ \frac{2 \kappa^2}{2\omega(x) + 3} \,
\left(2 V(x) - x \, \frac{dV}{dx} \right)
\,.
\end{array}
\right.
\eeq
This constitutes a projection of the trajectories on
the original two-dimensional constraint
surface in $(x, y, H)$ to the $(x,y)$ plane.
The projection yields two ``sheets'': the ``upper sheet''
marked by the $\mxp$ sign, and
the ``lower sheet'' marked by the the $\xpp$ sign in Eq. (\ref{V_dynsys}).

Standard procedure reveals that
the dynamical system (\ref{V_dynsys}) is endowed with two fixed points,
Table \ref{V_fp_eigenvalues} lists their conditions and eigenvalues.
The first fixed point $\Psi_{\bullet}$ satisfies \cite{burd_coley, faraoni3}
\be
\label{V_Psi_bullet}
\frac{\dd V}{\dd \Psi}\Big|_{\Psi_{\bullet}} \Psi_{\bullet} - 2
V(\Psi_{\bullet})=0 \, ,
\ee
which matches the second limit (\ref{second_GR_limit}),
discussed in Sec. 2.
The second fixed point $\Psi_{\star}$ satisfies
\be
\label{V_Psi_star}
\frac{1}{2\omega(\Psi_{\star}) +3} = 0 \,,
\qquad
\frac{1}{(2\omega(\Psi_{\star})+3)^2}\frac{\dd \omega}{\dd
\Psi}\Big|_{\Psi=\Psi_{\star}} \neq 0,
\ee
i.e. exactly the same conditions (a)-(d)
as the limit of general relativity for flat
FLRW STT cosmology, discussed in the end of Sec. 2.

\begin{table}[t] 
\begin{center}
\begin{tabular}{lclcl}
 & \quad & Condition & \quad & Eigenvalues
% \vspace{2mm}
\\ \hline \\
$\Psi_{\bullet} $ && 
$\frac{dV}{d\Psi}|_{\Psi_{\bullet}} \Psi_{\bullet} - 2 V(\Psi_{\bullet})=0$ &&
$\pmp \left[ -\sqrt{ \frac{3 \kappa^2 V}{4 \Psi}} \pm 
\sqrt{ \frac{\kappa^2}{2\omega+3} \left( (6\omega + 25) \frac{V}{4\Psi} - 2 \Psi \frac{d^2V}{d\Psi^2} \right) } \right]_{\Psi_{\bullet}}$  \vspace{2mm}\\
$\Psi_{\star} $ && 
$\frac{1}{2\omega(\Psi_{\star}) +3} = 0$, 
$\frac{1}{(2\omega(\Psi_{\star})+3)^2} \frac{\dd \omega}{\dd \Psi} \neq 0$ &&
$\pmp \left[ -\sqrt{ \frac{3 \kappa^2 V}{4 \Psi} } \pm 
\sqrt{ \frac{4 \kappa^2}{(2\omega +3)^2} \frac{d\omega}{d\Psi} \left(\Psi \frac{dV}{d\Psi} - 2V \right) + \frac{3 \kappa^2 V}{4 \Psi} } \right]_{\Psi_{\star}}$  
\vspace{2mm}\\
\hline
\end{tabular}
\end{center}
\caption{Fixed points and their eigenvalues for the $V \not\equiv 0$, $\rho \equiv 0$ case. 
}
\label{V_fp_eigenvalues}
\end{table}

From Eq. (\ref{H_constraint}) it is straightforward to compute that
the values of $H$
corresponding to the fixed points $\Psi_{\bullet}$ and $\Psi_{\star}$ are
$H_{\bullet}= \pmp 
\sqrt{ \frac{\kappa^2 V(\Psi_{\bullet})}{3
\Psi_{\bullet}}}$ and $
H_{\star}= 
\pmp \sqrt{ \frac{\kappa^2 V(\Psi_{\star})}{3 \Psi_{\star}}}$,
respectively. The result, which mimics
de Sitter evolution in general relativity, was expected, since we
saw in Sec. 2 that under the fixed point conditions (\ref{V_Psi_bullet})
and (\ref{V_Psi_star}) the full
STT equations (\ref{general_dynsys_x})-(\ref{general_dynsys_H}) reduce to
the equations of general relativity.

%%%%%%%%%%%%%%%%%%%%%%%%%%%%%%%%%%%%%%%%%%%%%%%%%%%%%%%%%%%%%%%%
%%%%%%%%%%%%%%%%%%%%%%%%%%%%%%%%%%%%%%%%%%%%%%%%%%%%%%%%%%%%%%

\section{Fixed points for matter domination
($V \equiv 0$, $\rho \not\equiv 0$)}

In the case of cosmological matter ($\rho>0$) and vanishing scalar potential
the Friedmann constraint restricts the solutions onto
a three-dimensional surface in four phase space dimensions
$(x \equiv \Psi, y \equiv \dot\Psi, H, \rho)$.
However, the system is amenable to a change of the time variable \cite{dn} 
that allows to combine the field equations into
a dynamical equation for
the scalar field which does not manifestly contain the scale factor or
matter density.
In the Jordan frame this amounts to defining a new time variable, 
$dp = h_c dt \equiv \left| H + \frac{\dot{\Psi}}{2\Psi} \right| dt \,$,
and deriving from Eqs. (\ref{00})--(\ref{deq})
a ``master'' equation for the scalar field \cite{sanq, meie2}.  
It can be written in a form of a dynamical
system for variables $\Psi \equiv x$, $\Psi' \equiv z$ 
($f' \equiv df/dp$)
\beq
%\left \{ \begin{array}{rcl}
\left \{ \begin{array}{rcl}
\! \! x' &= &z \\
\! \! z' &= & \! \!
\pmp {\frac {(2 \omega(x) +3) (1-\rw) }{ 8 x^2}} \, z^3
\! + \! \left( \frac{3 (1+\rw)}{4 x}
\! - \! \frac{1}{2 \omega(x) +3} \frac{d\omega(x)}{dx} \right)  z^2
\! \mpp \!  \frac{3 (1 - \rw) }{2}\, z
\! + \! \frac{3(1-3 \rw)}{(2\omega(x)+3)}\, x 
\label{j_master}
\,.
\end{array}
\right.
\eeq
The signs in Eq. (\ref{j_master}) correspond to 
the ``upper'' and the ``lower'' sheet 
as before in accordance
with the constraint equation (\ref{H_constraint}).

The ``master'' equation can be relied on as long as
$h_c= |H + \frac{\dot{\Psi}}{2\Psi}|$ is finite.
At $\Psi=0$ the quantity $h_c$ diverges making the $t$-time to stop
with respect to the $p$-time. Hence all $t$-time trajectories
with finite $\dot{\Psi}$  get mapped to $\Psi'=0$, giving
a false impression of a fixed point there.
However, in Sec. 2 we concluded that  $\Psi=0$ comes with a
space-time singularity and exclude it from present analysis.

The other problematic points can be discussed by noticing that
in terms of the new time variable $p$
the Friedmann constraint (\ref{00}) can be written as
\beq \label{jf_ptime_friedmann}
h_c^2= \frac{\kappa^2  \, \rho
}{3\Psi \left( 1-
\frac{(2\omega(\Psi)+3)\, \Psi'^2}{12 \Psi^2} \right)} \,.
\eeq
To keep $h_c$ real, the right hand side
of Eq. (\ref{jf_ptime_friedmann})
must be nonnegative, thus constraining the dynamically allowed regions of
the
two-dimensional phase space ($\Psi, \Psi'$) of the scalar field.
It is also important to verify that 
a fixed point in the $p$-time does indeed
correspond to a fixed point in the cosmological $t$-time.
But since on physical grounds it is reasonable to consider 
only the trajectories with finite $h_c$, it
is immediate that $\Psi'=0$ implies $\dot{\Psi}=0$ 
due to redefinition $p=p(t)$.

Let us consider the cosmological matter behaving like dust ($\rw=0$).
An argument completely analogous to the one put forth for Eq.
(\ref{taylor_expansion}), reveals a single fixed point,
satisfying the conditions (a)-(d)
dubbed as the limit of general relativity for flat FLRW STT.
The corresponding eigenvalues, to be evaluated at the fixed point
coordinate,
are given in Table \ref{matter_fp_eigenvalues}.
In particular, this point is an attractor on the ``upper'' sheet if
$\frac{\dd \omega}{\dd \Psi} >0$, while on the ``lower'' sheet attractor
behavior is not possible.
From Eq. (\ref{H_constraint}) now it also follows
that at the fixed point the evolution of the universe obeys the usual
Friedmann equation from general relativity,
$H_{\star}^2=\frac{\kappa^2 \rho}{3 \Psi_{\star}}$.
This is expected, as the fixed point conditions were identical to the
general relativity limit.

\begin{table}[t] 
\begin{center}
\begin{tabular}{lclclcl}
Case & \qquad & Fixed point & \qquad & Condition & \quad & Eigenvalues \vspace{2mm}\\
\hline \\
$\rw=0$&&
$\Psi_{\star}$ && $\frac{1}{2 \omega(\Psi_{\star})+3}=0$, 
$\frac{1}{(2\omega(\Psi_{\star})+3)^2} \frac{\dd \omega}{\dd \Psi} \neq 0$ && 
$\pmp \left[ -\frac{3}{4} \pm \frac{3}{4} 
\sqrt{1 - \frac{32}{3} \frac{\Psi}{(2 \omega+3)^2} \frac{d \omega}{d \Psi}}
\right]_{\Psi_{\star}}$ \vspace{2mm}\\
$\rw={1 \over 3}$ && none &&
\vspace{4mm}\\
\hline
\end{tabular}
\end{center}
\caption{Fixed points and their eigenvalues for the $\rho \not\equiv 0$,
$V \equiv 0$ case.}
\label{matter_fp_eigenvalues}
\end{table}

The dynamical system in the radiation dominated regime ($\rw=\frac{1}{3}$)
has no fixed points. For small values of $\Psi'$
the system is ruled by friction on the ``upper'' sheet, as the $\mxp$ sign
of the dominating term forces the vector flow to converge
to the $\Psi'' = 0$ axis.
On the ``lower'' sheet, the the effect is the opposite (anti-friction).

%%%%%%%%%%%%%%%%%%%%%%%%%%%%%%%%%%%%%%%%%%%%%%%%%%%%%%%%%%%%%%%%%%%%%%%%
%%%%%%%%%%%%%%%%%%%%%%%%%%%%%%%%%%%%%%%%%%%%%%%%%%%%%%%%%%%%%%%%%%%%%%

\section{Conclusion}

We have considered flat FLRW cosmological models in general
scalar-tensor theories with arbitrary coupling function $\omega(\Psi)$
and scalar potential $V(\Psi)$ in the Jordan frame. Using the methods
of dynamical systems we have 
%described the general geometry of the phase space and
found the scalar field fixed points in two distinct
asymptotic regimes: potential domination
($V\not\equiv 0, \rho \equiv 0$), and
matter domination ($V \equiv 0, \rho \not\equiv 0 $).
In nutshell there are two types of fixed points arising
from different mechanisms: $\Psi_{\bullet}$ from a condition on
the potential 
%(equalling the local extremum of the Einstein frame potential)
and $\Psi_{\star}$ from the singularity
of the scalar field kinetic term.
Approaching both types of fixed points
the cosmological equations coincide
with those of general relativity, yielding de Sitter expansion in
the potential domination case and Friedmann evolution in the matter
domination
case.
However, for the Solar System experiments in the PPN framework
only the fixed points of $\Psi_{\star}$ type give
predictions identical with those of general relativity.
The nature of fixed points (attractor or otherwise) depends
on the functional forms of
$\omega(\Psi)$ and $V(\Psi)$ according to corresponding
eigenvalues given in Tables 1 and 2.
Therefore, in Jordan frame analysis, general relativity is an attractor
for a large class of scalar-tensor models, but not for all.

Provided the transformation relating the Jordan and 
the Einstein frame is regular,
there is an exact correspondence between the two frames and the Jordan frame
phase space results should carry over to the Einstein 
frame \cite{meie3, dicke, flanagan, faraoni2}.
This is the case for the fixed point $\Psi_{\bullet}$.
However, as the transformation of the scalar field fails to be regular in
the limit of
general relativity, the properties of the $\Psi_{\star}$ fixed point may be
altered in the Einstein frame \cite{meie3}.
To establish whether or how the correspondence holds in this case calls for
a separate matching investigation in the Einstein frame.

%\medskip
\section*{ Acknowledgements}

This work was supported by the Estonian Science Foundation
Grants Nos. 7185, JD131 (MS) and by Estonian Ministry for Education and Science
Support Grant No. SF0180013s07.

\end{document}